\documentclass[twocolumn,showpacs,showkeys,reprint,amsmath,amssymb,aps]{revtex4-1}

\usepackage{graphicx}
\usepackage{dcolumn}
\usepackage{bm}
\usepackage{tabularx}
\usepackage{amsmath}

\begin{document}

\title{Ultralow-temperature heat transport in the quantum spin liquid candidate Ca$_{10}$Cr$_7$O$_{28}$ with bilayer kagome lattice}

\author{J. M. Ni,$^1$ Q. Y. Liu,$^2$ Y. J. Yu,$^1$ E. J. Cheng,$^1$ Y. Y. Huang,$^1$ Z. Y. Liu,$^2$ X. J. Wang,$^2$ Y. Sui,$^{2,\dag}$ and S. Y. Li$^{1,3,*}$}

\affiliation
{$^1$State Key Laboratory of Surface Physics, Department of Physics, and Laboratory of Advanced Materials, Fudan University, Shanghai 200433, China\\
 $^2$Department of Physics, Harbin Institute of Technology, Harbin 150001, China\\
 $^3$Collaborative Innovation Center of Advanced Microstructures, Nanjing 210093, China
}

\date{\today}

\begin{abstract}
Recently, a novel material with bilayer kagome lattice Ca$_{10}$Cr$_7$O$_{28}$ was proposed to be a gapless quantum spin liquid, due to the lack of long-range magnetic order and the observation of broad diffuse excitations. Here, we present the ultralow-temperature thermal conductivity measurements on single crystals of Ca$_{10}$Cr$_7$O$_{28}$ to detect its low-lying magnetic excitations. At finite temperatures, with increasing the magnetic fields, the thermal conductivity exhibits a clear dip around 6 T, which may correspond to a crossover in the magnetic ground state. At the zero-temperature limit, no residual linear term is found at any fields, indicating the absence of gapless itinerant fermionic excitations. Therefore, if the spinons do exist, they are either localized or gapped. In the gapped case, the fitting of our data gives a small gap $\Delta \sim$ 0.27(2) K. These results put strong constraints on the theoretical description of the ground state in this quantum spin liquid candidate.
\end{abstract}

\maketitle

Conventional magnets always develop long-range magnetic order at low temperatures where the phase transition can be described by a broken symmetry under Landau paradigm. In contrast, quantum spin liquids (QSLs) are such exotic states that the spins remain fluctuating and highly entangled with no magnetic order and spontaneous symmetry breaking even down to zero temperature \cite{R31,R5,R32,R11}. Interestingly, fractionalized excitations such as spinons with gauge fields can emerge from the QSL states \cite{R31,R32,R11,R5}. Since Anderson's seminal proposal in the 1970s \cite{T7}, a large number of experimental efforts have been made to search for QSL in real materials \cite{Y4,R27,R26,Y14,R13,Y16,Y8,Y10,R7,Y13,R24,Y23,R8,R9,CPL}, due to its unique physical properties, potential link with high-temperature superconductors \cite{T8}, and possible applications such as the topological quantum computation \cite{R36}. Many QSL candidates are proposed, among which the triangular-lattice organics $\kappa$-(BEDT-TTF)$_2$Cu$_2$(CN)$_3$ and EtMe$_3$Sb[Pd(dmit)$_2$]$_2$ \cite{Y4,R27,R26,Y14,R13,Y16} and kagome-lattice Herbersmithite ZnCu$_3$(OH)$_6$Cl$_2$ \cite{Y8,Y10,R7,Y13} are promising ones. However, no material has been widely accepted as a real QSL so far.

Recently, a new QSL candidate Ca$_{10}$Cr$_7$O$_{28}$ with kagome bilayers was discovered, stimulating a number of researches on this novel material \cite{R1,R2,R3,R37,R38,theory}. Ca$_{10}$Cr$_7$O$_{28}$ crystallizes in $R3c$ space group \cite{R1,R2}. There are seven chromium ions with different valences per formula unit, six of which are magnetic Cr$^{5+}$ while the remaining one is nonmagnetic Cr$^{6+}$, coordinated by oxygen tetrahedra \cite{R2}, which are shown in red and orange colors in Fig. 1(a), respectively. The magnetic Cr$^{5+}$ ions have spin $1/2$ moments, forming two inequivalent distorted kagome layers, stacked along the $c$ axis, as plotted in Fig. 1(b) \cite{R1,R2}. The kagome lattice is constructed by corner-sharing triangles, causing a large geometrical frustration. These two kagome layers are coupled together into a bilayer through the ferromagnetic interactions that connect the ferromagnetic triangles of one layer to the antiferromagnetic triangles in the other layer \cite{R1}. The ferromagnetic couplings are dominant among all the interactions, leading to a Hamiltonian with isotropic ferromagnetic interactions, which is rare for a QSL candidate \cite{R1}. Since the bilayers are isolated from each other, Ca$_{10}$Cr$_7$O$_{28}$ can be viewed as a two-dimensional bilayer kagome system \cite{R1}.

No magnetic order was observed by muon spin relaxation ($\mu$SR) measurements down to 0.019 K, far below the Curie-Weiss temperature $\theta$$_W$ = 2.35 K \cite{R1}. Spins are fluctuating coherently down to the lowest temperature, as revealed by $\mu$SR and a.c. susceptibility \cite{R1}. A broad and diffuse continuum was found in the inelastic neutron scattering (INS) spectrum which was argued to be the spinon excitation, a hallmark of QSL \cite{R1}. Combining the huge magnetic heat capacity down to 0.3 K \cite{R3} and the gapless neutron scattering spectrum \cite{R1}, Balz $et$ $al.$ strongly suggested Ca$_{10}$Cr$_7$O$_{28}$ be a gapless QSL \cite{R1,R3}. Furthermore, two crossovers were observed at $\mu_0H$ = 1 T and 6 T, based on the INS and magnetic specific heat measurements, showing a rich phase diagram in Ca$_{10}$Cr$_7$O$_{28}$ when applying magnetic fields \cite{R3}. A latest theoretical work using large-scale semiclassical molecular dynamics simulations argued that Ca$_{10}$Cr$_7$O$_{28}$ is a unique spin liquid due to its possibility that both slowly-fluctuating spiral spin liquid at low energies and fast-fluctuating U(1) spin liquid at high energies exist on multiple timescales \cite{theory}.

To establish the knowledge of ground states of QSL candidates, details of low-lying excitations would be one of the top priorities. It has been proven that the ultralow-temperature thermal conductivity measurement is a powerful technique to study the low-lying excitations in QSL candidates \cite{R13,R26,R4}. Taking the triangular-lattice organics $\kappa$-(BEDT-TTF)$_2$Cu$_2$(CN)$_3$ and EtMe$_3$Sb[Pd(dmit)$_2$]$_2$ as examples, thermal conductivity results imply a tiny gap \cite{R26} and gapless mobile fermionic excitations \cite{R13}, respectively. For the hotly-debated QSL candidate YbMgGaO$_4$, no magnetic thermal conductivity was detected \cite{R4}.

In this paper, we report the ultralow-temperature thermal conductivity measurements on single crystals of Ca$_{10}$Cr$_7$O$_{28}$. A dip around 6 T in the field dependence of thermal conductivity is clearly observed at finite temperatures, indicating a possible crossover in the magnetic ground state. More importantly, no residual linear term is detected, demonstrating the absence of gapless itinerant excitations with fermionic statistics. These results suggest that the magnetic excitations are either localized or gapped in this QSL candidate.

\begin{figure}
\includegraphics[clip,width=8cm]{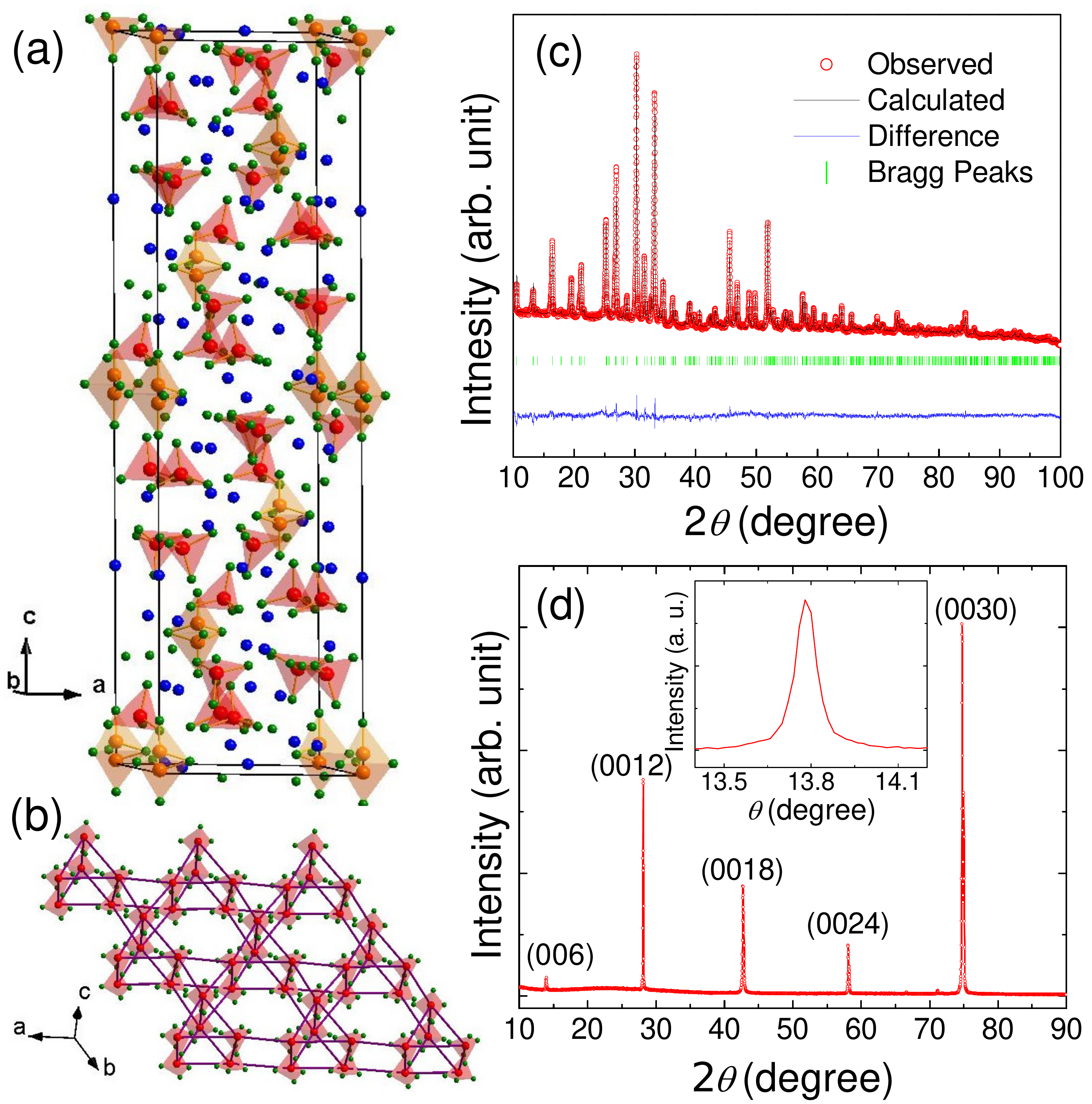}
\caption{(a) Crystal structure of Ca$_{10}$Cr$_7$O$_{28}$. O$^{2-}$, Ca$^{2+}$, Cr$^{5+}$ and Cr$^{6+}$ are labeled as green, blue, red and orange balls, respectively. Chromium ions are surrounded by four oxygen ions, forming CrO$_4$ tetrahedras which are shown by the same color as central chromium ions. (b) Schematic of distorted kagome bilayer showing only Cr$^{5+}$ and O$^{2-}$. The purple lines display the bilayer kagome network, and the colors for Cr$^{5+}$ and O$^{2-}$ are the same as in (a). (c) Rietveld refinement of the powder X-ray diffraction data for Ca$_{10}$Cr$_7$O$_{28}$. (d) Room-temperature XRD pattern from the largest surface of the Ca$_{10}$Cr$_7$O$_{28}$ single crystal. Only (00$l$) Bragg peaks are found. Inset: X-ray rocking scan curve of (0012) Bragg peak. The full width at half maximum of 0.09$^\circ$ indicates the high quality of the sample.}
\end{figure}

High-quality Ca$_{10}$Cr$_7$O$_{28}$ single crystals were grown by the traveling solvent technique in an optical floating zone furnace (IR Image Furnace G3, Quantum Design Japan) \cite{R1,R2}. A piece of single crystal was crushed to powder, and the Rietveld refinement of the powder X-ray diffraction (XRD) data confirmed the crystal structure of our samples, shown in Fig. 1(c). The single crystal orientation was determined by using Laue back-reflection, then a flat surface of (00$l$) planes was obtained by cutting the large crystal. The orientation of this surface was further confirmed by the XRD measurement, which is plotted in Fig. 1(d). The quality of Ca$_{10}$Cr$_7$O$_{28}$ single crystals was checked by the X-ray rocking scan, shown in the inset of Fig. 1(d). The full width at half maximum (FWHM) is only 0.09$^\circ$, indicating the high quality of the samples. Ca$_{10}$Cr$_7$O$_{28}$ single crystal for thermal conductivity measurements was cut and polished into a rectangular shape of dimensions 2.60 $\times$ 0.96 mm$^2$ in the $ab$ plane, with a thickness of 0.17 mm along the $c$ axis. The thermal conductivity was measured in a dilution refrigerator, using a standard four-wire steady-state method with two RuO$_2$ chip thermometers, calibrated $in$ $situ$ against a reference RuO$_2$ thermometer. Magnetic fields were applied along the $c$ axis.

\begin{figure}
\includegraphics[clip,width=6.6cm]{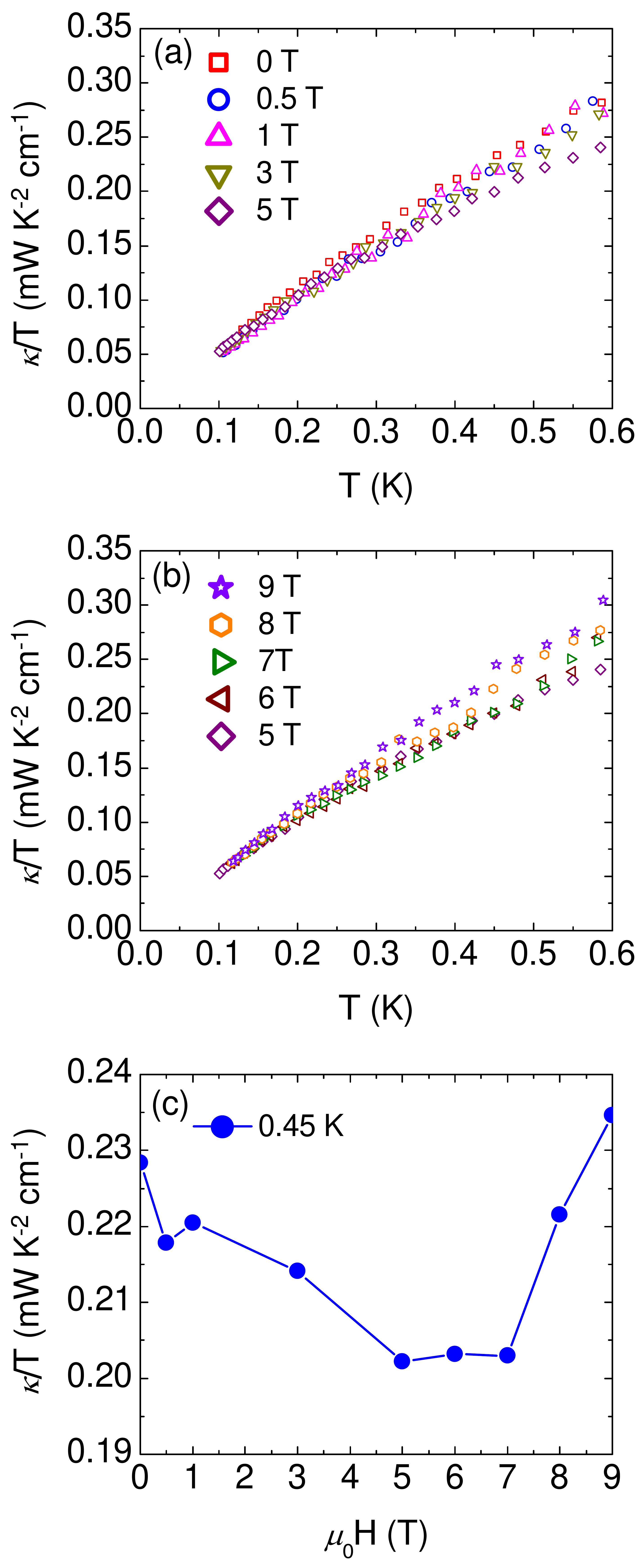}
\caption{(a) and (b) The in-plane thermal conductivity of Ca$_{10}$Cr$_7$O$_{28}$ single crystal at various magnetic fields. (c) Field dependence of the $\kappa/T$ at $T$ = 0.45 K. A dip around $\mu_0H$ = 6 T is clearly observed in the field dependence of $\kappa/T$ when increasing the magnetic fields.}
\end{figure}

Figure 2(a) and 2(b) present the in-plane thermal conductivity of Ca$_{10}$Cr$_7$O$_{28}$ single crystal in magnetic fields up to 9 T. The field-dependence of the $\kappa/T$ at $T$ = 0.45 K is also presented in Fig. 2(c). As the magnetic field increases, the $\kappa/T$ first drops until $\mu_0H$ $\sim$ 5 T. Then the thermal conductivity remains nearly the same between 5 T and 7 T, followed by a sharp increase. This dip around 6 T may correspond to the crossover suggested in Ref. \cite{R3} where a broad peak was found in the magnetic specific heat at $\mu_0H$ $\sim$ 6 T. We will come back to this point later.

\begin{figure}
\includegraphics[clip,width=6.6cm]{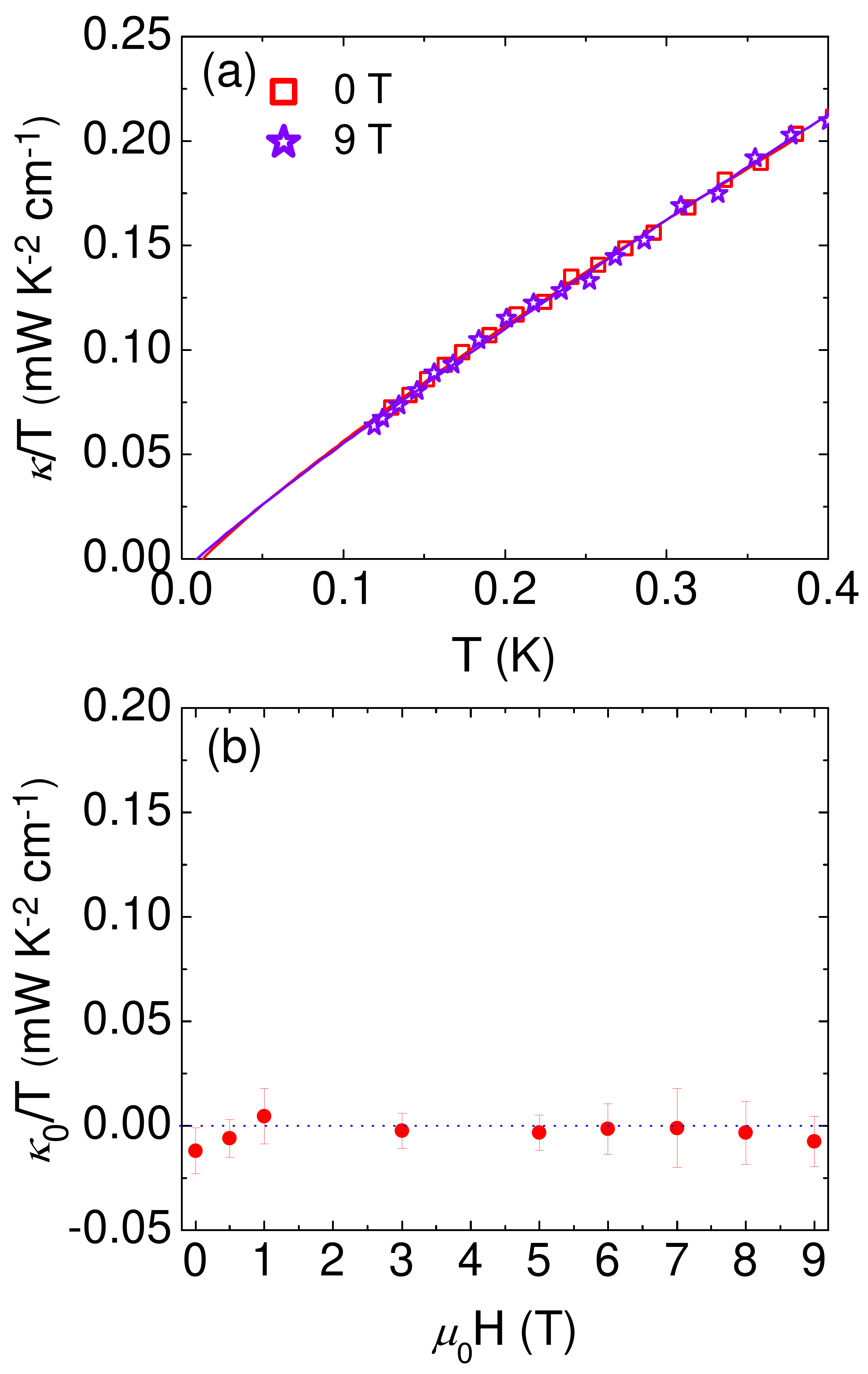}
\caption{(a) The in-plane thermal conductivity of Ca$_{10}$Cr$_7$O$_{28}$ single crystal at $\mu_0H$ = 0 T and 9 T, respectively. Solid lines are the fits of the thermal conductivity data to $\kappa/T$ = $a$ + $bT^{\alpha-1}$ at both magnetic fields below 0.4 K. (b) Field dependence of the residual linear term $\kappa_0/T$. These $\kappa_0/T$ values are all negligible within our experimental error bar at any magnetic fields we have measured.}
\end{figure}

In Fig. 3(a), we fit the thermal conductivity data below 0.4 K for $\mu_0H$ = 0 and 9 T to examine the residual linear term. In a solid, the contributions to thermal conductivity come from various quasiparticles, such as phonons, electrons, magnons, and spinons. Since Ca$_{10}$Cr$_7$O$_{28}$ is an insulator, electrons do not contribute to the thermal conductivity at such low temperatures at all. One also does not need to consider the magnon contribution due to the absence of magnetic order down to 0.019 K. Therefore, the ultralow-temperature thermal conductivity can be fitted by the formula $\kappa/T$ = $a$ + $bT^{\alpha-1}$, in which $aT$ and $bT^{\alpha}$ represent itinerant gapless fermionic magnetic excitations (if they do exist) and phonons respectively. The fitting of the 0 T and 9 T data below 0.4 K gives $a$ $\equiv$ $\kappa_0/T$ = -0.012 $\pm$ 0.011 mW K$^{-2}$ cm$^{-1}$, $\alpha$ = 1.78 $\pm$ 0.04, and $a$ = -0.007 $\pm$ 0.012 mW K$^{-2}$ cm$^{-1}$, $\alpha$ = 1.93 $\pm$ 0.05, respectively. Comparing with our experimental error bar $\pm$5 $\mu$W K$^{-2}$ cm$^{-1}$, the residual linear terms $\kappa_0/T$ of Ca$_{10}$Cr$_7$O$_{28}$ at both magnetic fields are virtually zero. Such negligible $\kappa_0/T$ is also obtained for the magnetic fields between 0 and 9 T, as plotted in Fig. 3(b).

The absence of $\kappa_0/T$ is reasonable for $\mu_0H$ = 9 T, because the magnetic excitations with a gap larger than 1 meV were observed at $\mu_0H$ = 9 T (close to the fully polarized state above 12 T) in the INS experiment \cite{R3}. As a result, the thermal conductivity at $\mu_0H$ = 9 T is purely contributed by phonons, without contributions from gapless fermionic excitations or magnons at such low temperatures. Therefore, the reduction of the thermal conductivity around 6 T may be explained in terms of the additional scattering of phonons. Indeed, a broad peak around $\mu_0H$ $\sim$ 6 T in the magnetic specific heat was observed \cite{R3}. It was considered as a signature of an crossover, which is linked to the changes in the magnetic excitations \cite{R3}. Interestingly, despite the fact that another crossover at $\mu_0H$ = 1 T was also suggested due to the distinct weak peak of magnetic specific heat and a new magnetic mode emerging in the INS measurements \cite{R3}, we do not find an anomaly in our ultralow-temperature thermal conductivity at this field.

Now we would like to focus our attention to the data at $\mu_0H$ = 0 T and discuss the implication of our ultralow-temperature thermal conductivity results. Generally, there are two types of QSLs according to whether they exhibit an excitation gap or not \cite{R5}. For Ca$_{10}$Cr$_7$O$_{28}$ at $\mu_0H$ = 0 T, two distinct bands of magnetic excitations with energy ranges 0-0.6 meV and 0.7-1.5 meV were found by INS experiments \cite{R1}. These broad and diffuse excitations were argued to be the fractionalized spinon continuum rather than the sharp and dispersive spin-wave excitations \cite{R1}. Furthermore, a huge magnetic heat capacity down to 0.3 K suggest a large spin gap is unlikely in Ca$_{10}$Cr$_7$O$_{28}$ at zero field \cite{R3}. Therefore, one may expect a gapless scenario for Ca$_{10}$Cr$_7$O$_{28}$, such as a spin liquid with a spinon Fermi surface or a Dirac spin liquid with a nodal fermionic spinon at the Dirac point. Recently, A gapless spiral spin liquid is suggested for Ca$_{10}$Cr$_7$O$_{28}$ at low energies using large-scale semiclassical molecular dynamics simulations \cite{theory}. Under all these circumstances, a non-zero residual linear term $\kappa_0/T$ should be found in the ultralow-temperature thermal conductivity measurements \cite{R14,R17,R15} due to the contribution to $\kappa$ from fermionic magnetic excitations. In this context, it is surprising that we do not observe any significant magnetic contribution to the thermal conductivity of Ca$_{10}$Cr$_7$O$_{28}$. By contrast, the gapless QSL candidate EtMe$_3$Sb[Pd(dmit)$_2$]$_2$ has a value of $\kappa_0/T$ as large as 2 mW K$^{-2}$ cm$^{-1}$ \cite{R13}. The negligible $\kappa_0/T$ in Ca$_{10}$Cr$_7$O$_{28}$ indicates that either the gapless spinons do exist but do not conduct heat for some reason or the excitations are gapped with a small gap which is masked by the nuclear and incoherent scattering in the INS measurements below 0.2 meV \cite{R1}.

In the case that gapless spinons do not conduct heat, one possible mechanism is localization. This speculation is supported by the estimation of the mean free path of spinons according to the kinetic formula $\kappa$ = $\frac{1}{3}$${C_m}$$v_s$$l_s$, where $C_m$, $v_s$ and $l_s$ are the magnetic heat capacity, velocity of spinons and mean free path of spinons, respectively. The velocity $v_s$ is estimated as $Ja/\hbar$ $\approx$ 1.6 $\times$ 10$^3$ $m/s$ ($a$ = 10.8 \AA\ is the inter-spin distance and $J$ $\approx$ 1 meV is the value of the exchange interaction) and $C_m$ is about 4.2 J K$^{-1}$ mol$^{-1}$ at 0.3 K \cite{R1,R2,R13,R18}. Even if we assume that $\kappa$ at 0.3 K is totally contributed by spinons, the $l_s$ would only be 17 \AA, less than two inter-spin distances. In contrast, the gapless magnetic excitations in EtMe$_3$Sb[Pd(dmit)$_2$]$_2$ can travel through about 1000 inter-spin distances without being scattered \cite{R13}. The localization can be ascribed to the lattice disorder \cite{R4}. In fact, such lattice disorder is indeed an obstacle to the study of true ground state in various QSL candidates \cite{R19,R20,R21,R22,R23,R24}. Here for Ca$_{10}$Cr$_7$O$_{28}$, there are two inequivalent position Cr3A and Cr3B for Cr$^{6+}$ around the 6a(0,0,z) site, as can be seen from the plane-sharing tetrahedra in orange color in Fig. 1(a), where the up and bottom tetrahedra are Cr3BO$_4$ and Cr3AO$_4$ respectively \cite{R2}. Only one of the tetrahedra can be occupied \cite{R2}. The disorder among the Cr3A and Cr3B positions was reported in Ref. \cite{R2}, which may be the reason for the localization of spinon excitations.

\begin{figure}
\includegraphics[clip,width=6.6cm]{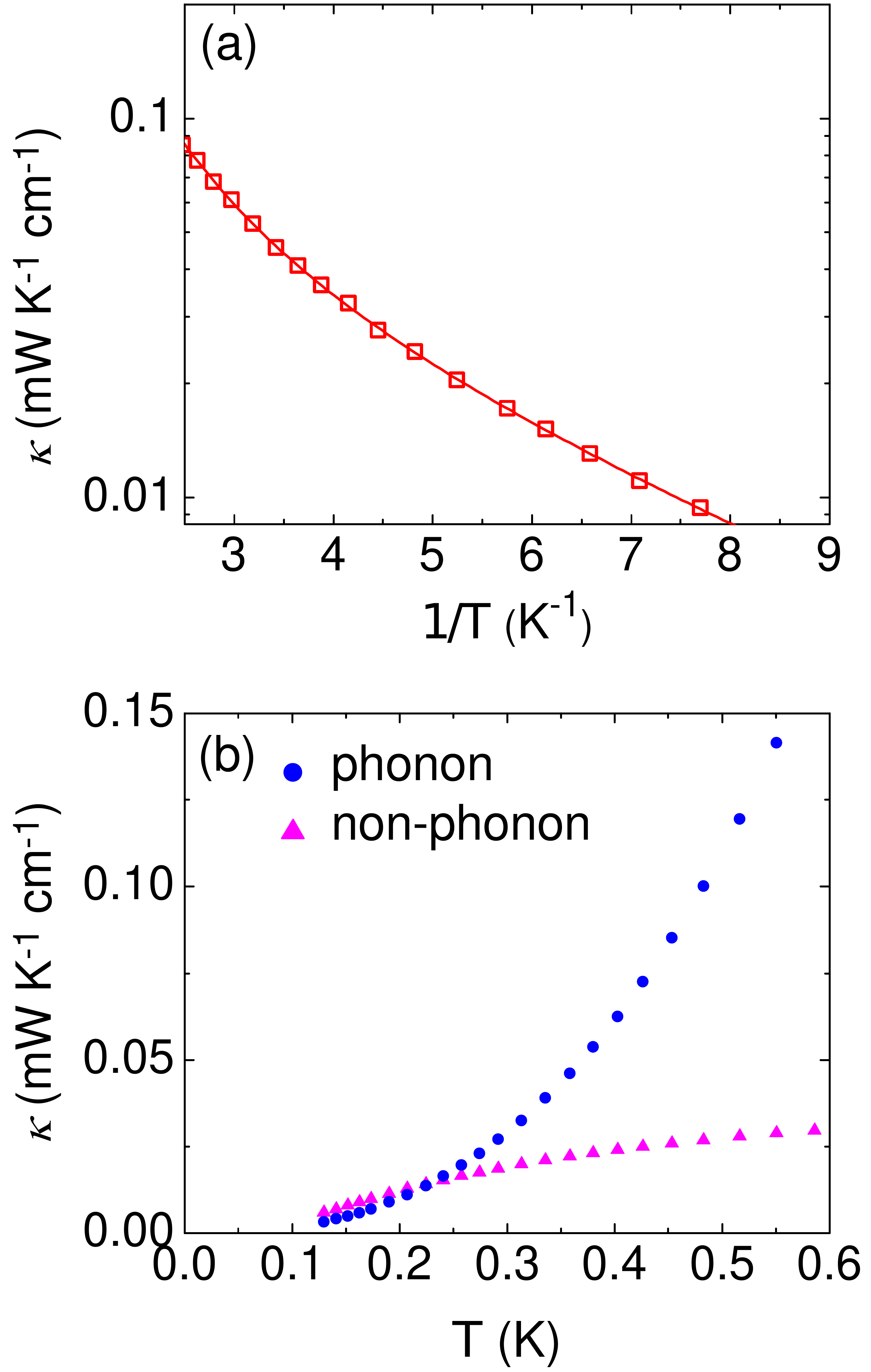}
\caption{(a) An Arrhenius plot of the thermal conductivity at $\mu_0H$ = 0 T. The solid line is the fit of the thermal conductivity to $\kappa$ = $A$$e^{-\Delta/k_BT}$ + $B$$T^\beta$ with $\Delta$ = 0.27 $\pm$ 0.02 K and $\beta$ = 2.6 $\pm$ 0.3. (b) Phonon part and non-phonon part of the thermal conductivity at $\mu_0H$ = 0 T calculated by the formula $\kappa_{ph}$ = $B$$T^\beta$ and $\kappa_{nph}$ = $A$$e^{-\Delta/k_BT}$, respectively. The parameters are obtained from the above fitting result.}
\end{figure}

A gapped QSL is another possible scenario to interpret our experimental results. As a similar case, a negligible $\kappa_0/T$ was also reported in the triangular-lattice QSL candidate $\kappa$-(BEDT-TTF)$_2$Cu$_2$(CN)$_3$ in spite of the large density of low-energy states manifested by a linear term of 15 mJ K$^{-2}$ mol$^{-1}$ in specific heat \cite{R26,R27}. The thermal conductivity of $\kappa$-(BEDT-TTF)$_2$Cu$_2$(CN)$_3$ can be fitted by $\kappa$ = $A$$e^{-\Delta/k_BT}$ + $B$$T^3$ \cite{R26}. The first term has an exponential temperature dependence, instead of $aT$ used in gapless situations, and the second term is the phonon part. The value of $\Delta$ $\approx$ 0.46 K is obtained, indicating a gapped QSL with a tiny gap compared to the exchange coupling $J$ $\sim$ 250 K \cite{R26}. Following the same procedure, we fit the thermal conductivity data of Ca$_{10}$Cr$_7$O$_{28}$ at 0 T and get the value of $\Delta$ = 0.29 $\pm$ 0.01 K. Since the power in the phonon part is typically between 2 and 3 due to the specular reflections of phonons at the sample surface \cite{R28,R29}, it is more reasonable to fit the data by $\kappa$ = $A$$e^{-\Delta/k_BT}$ + $B$$T^\beta$ (see Fig. 4(a)). As a result, $\beta$ = 2.6 $\pm$ 0.3 and $\Delta$ = 0.27 $\pm$ 0.02 K are obtained. Such a tiny gap can be masked by the high intensity of nuclear and incoherent scattering below 0.2 meV in the INS experiments \cite{R1}. This may account for the discrepancy between the gapless spectrum \cite{R1,R3} and our negligible $\kappa_0/T$. As can be seen in Fig. 4(b), the power-law phonon part of the thermal conductivity is of the same order as the exponential non-phonon part. This may explain the power $\alpha$ abnormally lower than 2 when we fit the total thermal conductivity, the sum of the two parts, to $\kappa/T$ = $a$ + $bT^{\alpha-1}$ (see Fig. 3(a)). The gap observed in $\kappa$-(BEDT-TTF)$_2$Cu$_2$(CN)$_3$ is interpreted as a vison gap \cite{R30}. However, due to the lack of appropriate theoretical models for the bilayer kagome QSL candidate Ca$_{10}$Cr$_7$O$_{28}$, it is hasty to jump to a conclusion what the essence of the small gap is.

Finally, a relatively trivial scenario that Ca$_{10}$Cr$_7$O$_{28}$ is a spin glass is also consistent with our ultralow-temperature thermal conductivity measurement. In fact, the a.c. susceptibility shows a broad maximum at about 0.33 K which shifts by 0.015 K from 158 Hz to 20.0 kHz \cite{R1}. The frequency-dependent peak evidencing a broad distribution of relaxation time is a typical hallmark for a spin glass \cite{SG}. Recently, the similar behavior was also observed in YbMgGaO$_4$  and YbZnGaO$_4$, which was argued as an evidence for the spin glass ground state \cite{YZGO}. More experimental and theoretical efforts are desired to determine the true ground state and excitations of Ca$_{10}$Cr$_7$O$_{28}$ by a combination of various techniques.

In summary, we have measured the ultralow-temperature thermal conductivity of Ca$_{10}$Cr$_7$O$_{28}$ single crystals. At finite temperatures, the field dependence of the thermal conductivity exhibits a dip around 6 T, corresponding to a crossover in the magnetic ground state. At zero-temperature limit, no residual linear term is observed at all fields, suggesting the absence of itinerant fermionic magnetic excitations in Ca$_{10}$Cr$_7$O$_{28}$. If the spinons do exist, they should be localized or be gapped with a small gap of $\Delta$ $\sim$ 0.27 $\pm$ 0.02 K. These results put strong constraints on the ground state and theoretical description of Ca$_{10}$Cr$_7$O$_{28}$.

We thank Y. Xu for helpful discussions. This work is supported by the Ministry of Science and Technology of China (Grant No: 2015CB921401 and 2016YFA0300503), the Natural Science Foundation of China, the NSAF (Grant No: U1630248), the Program for Professor of Special Appointment (Eastern Scholar) at Shanghai Institutions of Higher Learning, and STCSM of China (No. 15XD1500200). The work in Harbin Institute of Technology is supported by the Natural Science Foundation of China (No: 51472064).\\

\noindent $^\dag$ E-mail: suiyu$@$hit.edu.cn\\
\noindent $^*$ E-mail: shiyan$\_$li$@$fudan.edu.cn

\end{document}